\documentclass[reprint,prd, superscriptaddress, tightenlines, longbibliography, nofootinbib, eqsecnum, amsfonts, amsmath, floatfix, notitlepage]{revtex4-1}
\pdfoutput=1

\usepackage[utf8]{inputenc}
\usepackage{mathrsfs}
\usepackage{euscript}
\usepackage{epsfig}
\usepackage{graphics}
\usepackage{graphicx}
\usepackage{comment}
\usepackage{amsmath}
\usepackage{amssymb}
\usepackage{bm}
\usepackage[usenames,dvipsnames,svgnames,table]{xcolor}
\usepackage{xspace}
\usepackage{wasysym}
\usepackage{times}
\usepackage{appendix}
\usepackage{lipsum}
\usepackage[nolist,nohyperlinks]{acronym}
\usepackage{float}
\usepackage{simplewick}
\usepackage{natbib, ifthen}
\usepackage{hyperref}
\usepackage{longtable}
\usepackage[T1]{fontenc}
\newcommand{\isdraft}[1]{}
\newcommand{\vx}[0]{{\boldsymbol{x}}}
\newcommand{\vy}[0]{{\boldsymbol{y}}}
\renewcommand{\vr}[0]{{\boldsymbol{r}}}

\providecommand{\CAMB}{{\tt camb}}

\newcommand{\lag}{\left \langle}
\newcommand{\rag}{\right \rangle}
\newcommand{\av}[1]{\lag #1 \rag}

\newcommand{\beq}{\begin{equation}}
\newcommand{\enq}{\end{equation}}

\newcommand{\Covi}{{\textrm{Cov}^{-1}_{\phi}}}
\newcommand{\Cov}{{\textrm{Cov}_{\phi}}}
\newcommand{\Tr}{\textrm{Tr}}
\newcommand{\bL}[0]{ {\boldsymbol{L}} }
\newcommand{\bl}[0]{ {\boldsymbol{l}} }

\begin{document}
\newcommand{\Sussex}{Department of Physics \& Astronomy, University of Sussex, Brighton BN1 9QH, UK}
\author{Julien Carron}
\affiliation{\Sussex}
\email{j.carron@sussex.ac.uk}


\title{Optimal constraints on  primordial gravitational waves from the lensed CMB}
\date{\today}
\begin{abstract}
We demonstrate how to obtain optimal constraints on a primordial gravitational wave component in lensed Cosmic Microwave Background (CMB) data under ideal conditions. We first derive an estimator of the tensor-to-scalar ratio, $r$, by using an error-controlled close approximation to the exact posterior, under the assumption of Gaussian primordial CMB and lensing deflection potential. This combines fast internal iterative lensing reconstruction with optimal recovery of the unlensed CMB. We evaluate its performance on simulated low-noise polarization data targeted at the recombination peak. We carefully demonstrate our $r$-posterior estimate is optimal and shows no significant bias, making it the most powerful estimator of primordial gravitational waves from the CMB. We compare these constraints to those obtained from $B$-mode band-power likelihood analyses on the same simulated data, before and after map-level quadratic estimator delensing, and iterative delensing. Internally, iteratively delensed band powers are only slightly less powerful on average (by less than 10\%), promising close-to-optimal constraints from a stage IV CMB experiment.
\end{abstract}

\maketitle

\newcommand{\Stdat}{X^{\rm dat}}
\newcommand{\Stdatt}{X^{\rm dat,\dagger}}
\newcommand{\fsky}{f_{\rm sky}}
\newcommand{\planck}[0]{\textit{Planck}}
\section{Motivation}
After the completion of the \planck\ Cosmic Microwave Background (CMB) mission\cite{Akrami:2018vks}, the major target of the CMB community has now become precise measurement of the CMB polarization. The magnetic (B) part of the CMB polarization~\cite{Kamionkowski:1996ks, Zaldarriaga:1996xe} on degree scales is a unique signature of the stochastic background of primordial gravitational waves produced during inflation~\cite{Seljak:1996gy, Kamionkowski:2015yta}. Constraints on the tensor-to-scalar power spectrum ratio, $r$, are expected to increase by two orders of magnitude in precision within the next decade: CMB Stage IV (CMB-S4\footnote{\url{https://cmb-s4.org}}) has forecast sensitivity down to $r\sim 5\cdot 10^{-4}$, using delensed $B$-mode band powers after foreground cleaning \cite{Abazajian:2016yjj}. Lensing of the CMB photons by large-scale structures generates $B$ polarization that effectively appears as an  approximately white cosmic variance noise~\cite{Zaldarriaga:1998ar, Lewis:2006fu}. In order to reach such tight constraints on the primordial signal, successful delensing of this $5 \mu$K-arcmin noise is mandatory. Currently and for the next few years, the most faithful lensing tracer at the scales relevant for $B$-mode delensing is the Cosmic Infrared Background(CIB) \cite{Sherwin:2015baa}, able to achieve 40\% delensing on 60\% of the sky \cite{Aghanim:2018oex}, and possibly more in areas that are clean from galactic dust.  It is also possible to combine the CIB with other large-scale structure tracers \cite{Manzotti:2017oby, Yu:2017djs}, or with the CMB internal reconstruction \cite{Aghanim:2018oex}, in order to increase its fidelity to the CMB lensing field. Lensing estimates for CMB-S4 will be dominated by the internal reconstruction, using polarization quadratic estimators \cite{Okamoto:2003zw} or more powerful iterative estimators, first introduced by Ref.~\cite{Hirata:2003ka}. At the low instrumental noise levels expected for CMB-S4, iterative internal estimation from CMB polarization has been demonstrated on simulated data to give lensing reconstructions that are more than 90\% cross-correlated to the true lensing \cite{Seljak:2003pn,Carron:2017mqf}.

One may ask whether it could be possible, at least in principle, to do even better than these forecasts. The lensing deflections introduce  non-Gaussianities in the form of higher order statistics in the CMB temperature and polarization \cite{Lewis:2006fu}, which are used to reconstruct the lensing signal \cite{Hanson:2009kr}. Delensing will remove part of the non-Gaussianity, but only imperfectly, and some amount of information must remain beyond the power spectra. Hence, it is plausible that there may be room for alternative statistics that compress more information than delensed $B$-mode band-powers.

This paper has two main purposes. The first is to demonstrate how to obtain directly the posterior probability density (PDF) for $r$, from lensed CMB data. The posterior contains all the information on $r$, and constraints based on it are optimal. The second is to compare this optimal method to band-power likelihood analysis. Finding a posterior width in agreement with naive expectations will confirm current forecasting methods and our understanding of how well CMB experiments can constrain primordial gravitational waves.

Our approach uses an approximate, analytic marginalization of the large-scale structure lensing to build the statistics of interest, here the tensor-to-scalar ratio, $r$. This analytic marginalization is a fairly natural choice, and has been used already in a different context by Ref.~\cite{Hirata:2003ka}, where the aim was to obtain an optimal estimator of the lensing spectrum. In this paper we generalize the analytic marginalization to an error-controlled variational approximation and provide a rigorous discussion of its accuracy, showing that corrections are negligible for our purposes and the experimental configurations investigated. Variational principles have a long history in cosmology, at least dating back to J. Peebles minimum action principle to reconstruct large scale motions\cite{peebles1989tracing}.

We work in the flat-sky approximation. All our simulations use the \planck\ 2015 cosmology~\cite{Ade:2015xua}, with power spectra generated with the \CAMB~\citep{Lewis:1999bs} software. We use square maps of area $645\deg^2$, assuming periodic boundary conditions for simplicity, with pixels of $1.5$ arcmin on a side. Foreground cleaning is a major challenge to the quest for primordial gravitational waves. We do not consider these complications in this paper, assuming throughout we are working with foreground-cleaned maps with white noise power spectra.  We always use white Gaussian noise levels of $\sqrt{2} \cdot 1.5 \mu$K-arcmin in polarization, a instrument beam of 3 arcminutes, and consider CMB multipoles below $\ell = 3000$ only. The minimum multipole we probe in our flat-sky implementation of the sky patches is $\ell_{\rm min} = 14$, excluding the reionization peak. We consider 3 different levels of tensor modes, with tensor-to-scalar ratio $r_{\rm in}$ defined at the pivot scale $k = 0.05/\rm{Mpc}$ and a vanishing tensor spectral index. The first has $r_{\rm in} = 0.05$, close to current constraints $r_{0.002} < 0.062$ (95\% c.f.) from BICEP2/KECK array BK15 data in combination with \planck~\cite{Ade:2018gkx, Akrami:2018odb}. For the configuration just described, the nominal band powers are enough for a strong detection. Second, $r_{\rm in} = 0.01$, in which case the nominal band-powers cannot detect the waves decisively but the delensed band-powers can. Third, a vanishing amplitude $r_{\rm in} = 0.0$ where in all cases only upper limits can be placed from the data. 

The paper is built as follows. Sec.~\ref{sec:posterior} describes our approximation scheme to the exact posterior density function, and Sec.~\ref{sec:evaluation} gives details on our numerical implementation. In Sec.~\ref{sec:bandpowers}, we discuss our nominal and delensed band powers likelihood and implementation. We present in Sec.~\ref{sec:results} our results and summarise and conclude in Sec.~\ref{sec:summary}. One appendix details a couple of technical points for completeness.

\section{Posterior for tensor-to-scalar ratio}\label{sec:posterior}

Given CMB Stokes parameter polarization data $\Stdat = (Q^{\rm dat}, U^{\rm dat})$, and for a uniform prior on $r$, the posterior is proportional to the likelihood
\beq
p(r|\Stdat) \propto p(\Stdat|r).
\enq
Owing to the lensing by large scale structures, the CMB probability density function on the right-hand side is non-Gaussian in $X^{\rm dat}$ and does not have a simple analytical description. We may write, however, marginalising over possible lensing deflection maps,
\beq
\label{eq:likelihood}
p(\Stdat|r) = \int \mathcal D \phi \:p_\phi[\phi] p(\Stdat |\phi,r),
\enq
where $p_\phi$ is the probability density of the lensing potential, defined by the large-scale structure evolution. When the lensing map is known, the lensed CMB is still Gaussian, since the deflections just remap points on the sky to very good approximation~\cite{Lewis:2017ans}. Hence we can consider $p(\Stdat |\phi,r)$ Gaussian in $\Stdat$, with covariance determined by the lensed spectra of the CMB together with the specified amount of tensor mode, the transfer function and noise covariance matrix. Neglecting for simplicity the tiny cross-correlation between $\phi$ and the CMB $E$-polarization~\cite{Lewis:2011fk}, which is too small to impact our results\footnote{We do include properly all cross-correlations including $C_\ell^{\phi E}$ in all our simulations.}, we can write explicitly,
\beq \label{eq:cov}
\ln p(\Stdat |\phi,r) = -\frac 12 \Stdatt \rm{Cov}_\phi^{-1} \Stdat - \frac 12 \ln \det \rm{Cov}_{\phi},
\enq
with pixel-space covariance matrix
\beq
\Cov =\mathcal B D_\phi C^{\rm{unl, fid}}D_\phi^\dagger \mathcal B^\dagger + N.
\enq
Here, $\mathcal B$ is the  transfer function including instrument beam, $D_\phi$ is the lensing deflection operator that maps the unlensed CMB Stokes parameters to the ones deflected by $\nabla \phi$, $C^{\rm unl,  fid}$ is our set of \textit{unlensed} fiducial CMB spectra, and $N$ is the noise matrix. In position space and on the flat-sky, $D_\phi$ has the explicit representation $(D_\phi X)(\vx) \equiv X(\vx + \nabla \phi(\vx))$.

The model specified by Eq.~\eqref{eq:cov} neglects the very small effect of lensing of the polarization generated from reionization,  so that a single, common lensing deflection $D_\phi$ can be used. The CMB spectra contains the dependence on tensor modes
\beq
C_\ell^{\rm{unl, fid}} \equiv C_\ell^{\rm{scal, fid}} + r \: C_\ell^{\rm tens., fid}
\enq
We pick the prior on lensing maps $p_\phi$ in Eq.~\ref{eq:likelihood} to be Gaussian with power $C_L^{\phi\phi, \rm fid}$, calculated from the non-linear matter power\footnote{We follow standard practice of denoting lensing multipoles with $L$ and CMB multipoles with $\ell$.}. Owing to non-linear evolution in the late Universe, $p_\phi$ is not exactly Gaussian. However, non-linear effects are weak at scales relevant for the lensing $B$-modes (the non-linear contribution to the degree-scale lensing $B$-power is only a few percent), and generally the effects of non-Gaussian lenses is not expected to bias the lensing reconstruction from polarization \cite{Beck:2018wud, Bohm:2018omn}. Hence this choice is unlikely to significantly bias our results. In practice this assumption could simply be explicitly tested performing the reconstruction of this paper using lensing maps from realistic $N$-body simulations.

The integrand in Eq.~\ref{eq:likelihood} peaks at the most probable lensing map, given the data and candidate value $r$ for the tensor amplitude. Let $\phi^*(\Stdat,r)$ be this most probable potential map. We may then write
\beq
e^{-S[\phi]} \equiv p_\phi[\phi]\:p(X^{\rm dat} |\phi,r)
\enq
and expand the action $S[\phi]$ to second order around $\phi^*$,
\beq \label{eq:Sexp}
S[\phi] \sim S[\phi^*(r)]+ \frac 12 \left(\phi - \phi^*(r)\right) H_r[\phi^*(r)] \left(\phi - \phi^*(r)\right).
\enq
Both the lensing potential $\phi^*$ and the curvature $H_r$ at this point have explicit dependence on the data map, and explicit (through the CMB spectra) and implicit (through $\phi^*(r)$) dependence on $r$. In the following we suppress the data dependence for notational clarity. 

The lensing map marginalization, Eq.~\ref{eq:likelihood}, becomes trivial under this approximation. The result is, up to terms independent from $r$,
\beq\label{eq:posterior}
\ln p(r |X^{\rm dat}) \simeq  -S[\phi^*(r)] - \frac 12 \ln \det H_r[\phi^*(r)],
\enq
which forms the basis of our investigations. Leading corrections or alternative expansions are discussed below in Sec.~\ref{sec:corrections}.

In the remainder of this subsection we just note how $S[\phi^*]$ is related to the $B$-mode power, by looking at its dependence on $r$. Let us introduce the Wiener-filtered CMB maps $X^{\rm WF} = (E^{\rm WF}, B^{\rm WF})$ through
\beq \label{eq:XWF}
X^{\rm WF} \equiv C^{\rm unl, fid} D_\phi^\dagger \mathcal B^\dagger \Covi \Stdat.
\enq
These are the most probable unlensed CMB maps given the observed Stokes data and given the fiducial model entering the covariance $\Cov$. In particular, since the scalar unlensed $B$ power vanishes, $B^{\rm WF}$ is the most probable (maximum a posteriori) map of the $B$ tensor modes, assuming the fiducial value of $r$ and lensing map $\phi$ are the truth. By definition
\beq
\left.\frac{\delta S[\phi]}{\delta \phi}\right|_{\phi = \phi^*(r)} = 0,
\enq
hence $dS /dr = \partial S /\partial r$ for this lensing map. For this reason, the prior term does not contribute to the action $r$-derivative, only the likelihood defined in Eq.~\eqref{eq:cov}. It then follows (neglecting the tensor $E$ contribution)
\beq \label{eq:dSdr}
\begin{split}
&	\frac{dS[\phi^*(r)]}{d \ln r}= \frac 12 \sum_{\boldsymbol{\ell}}\frac{ \left|B_{\boldsymbol{\ell}}^{\rm WF}(r)\right|^2 -\av{\left|B_{\boldsymbol{\ell}}^{\rm WF}(r)\right|^2}}{r\:C_\ell^{BB,{\rm tens}}}.
\end{split}
\enq
The first term on the right-hand side comes from the quadratic part in the CMB likelihood. The sum $\sum_\bl$ runs over all 2D frequencies of our sky patch. The second term (the average of the first over data realizations) comes from the log-determinant $\ln \det \Cov$ $r$-derivative\footnote{This term is most easily derived realizing that on average, the likelihood variation vanishes, $\av{\partial_r \ln p(X^{\rm dat}|\phi, r)}_{X^{\rm dat}} = \partial_r\av{p(X^{\rm dat}|\phi, r)}_{X^{\rm dat}} = 0$, since $p$ is a properly normalized probability density for any value of $r$. Hence Eq.~\ref{eq:dSdr} must average to zero.}.

All dependence on the lensing map prior in the reconstructed $r$-posterior Eq.~\eqref{eq:dSdr} is absorbed into the lensing map reconstruction $\phi^*$. If the lensing map were exactly known, $B^{\rm WF}$ is the properly delensed $B$-mode map, there is no marginalization over the lensing map and Eq.~\eqref{eq:dSdr} directly gives the (derivative of log)$r$-posterior by trivial comparison to the expected power.

\subsection{Corrections and alternative approximations}\label{sec:corrections}

Eq.~\eqref{eq:posterior} is an approximation of the posterior PDF, that relies on the Gaussianity of the reconstructed lensing map. Lens reconstruction from the CMB is non-linear in the data, and some non-Gaussian features are expected, and visible, in standard lens reconstructions with perfectly Gaussian true input lensing\cite{Liu:2016nfs}. If the approximation is not good enough, resulting constraints might be biased, or suboptimal, possibly both. Hence it is very useful to be able to assess the size of the leading corrections. We discuss now how corrections can be evaluated, also giving us alternative expansion schemes. Sec.~\ref{sec:corrections_calc} later on demonstrates using these tools that Eq.~\eqref{eq:posterior} is accurate and unbiased.

We can proceed as follows. The exact posterior in Eq.~\eqref{eq:likelihood} integrates $e^{-S[\phi]}$. Using an arbitrary trial action $S_t$ as an approximation to $S$, we may write the exact identity
\beq\label{eq:approx}
\int \mathcal D \phi \:e^{-S[\phi]} = \left(\int \mathcal D\phi\: e^{-S_t[\phi]}\right) \av{ e^{-\Delta S[\phi]}}_{t},
\enq
where the average is with respect to probability density $e^{-S_t}$, and $\Delta S = S - S_t$ is the mismatch between the true and trial actions. After taking the logarithm, and for $S_t$ the natural expansion Eq.~\eqref{eq:Sexp}, the first term on the right-hand side (the normalization factor of the density $e^{-S_t}$) is our approximation Eq.~\eqref{eq:posterior}, and the second term encapsulates all errors. The point is that whenever $S_t$ is chosen quadratic in $\phi$, this error term is an average over Gaussian lensing maps. We can simulate these maps, and estimate this term by averaging over simulations. Unless the approximation is extremely good, in practice we can never probe directly the exponential with a reasonable number of Monte-Carlo's, but we obtain in this paper leading contributions. Using a standard cumulant expansion we can write the asymptotic expansion
\beq \label{eq:cumulants}
\ln \av{ e^{-\Delta S}}_t\sim \av{-\Delta S}_t + \frac 12 \left( \av{(\Delta S)^2}_t - \av{\Delta S}_t^2\right) +  ...
\enq
In particular, all expansions of the form
\beq
S_t[\phi] = S[\phi^*] + \frac 12 \left(\phi - \phi^* \right)^\dagger H\left(\phi - \phi^* \right),
\enq
for an arbitrary $H$ matrix results in (including here only the leading cumulant corrections)
\beq\begin{split}
&\ln p(r |X^{\rm dat}) \sim -S[\phi^*] - \frac 12 \ln \det H - \kappa_1 + \frac 12 \kappa_2,
\end{split}
\enq
where $\kappa_{1}$ and $\kappa_2$ are the mean and variance of $\Delta S$ from a Gaussian ensemble of lensing maps with inverse covariance $H$. This provides alternative expansion schemes and useful consistency checks, where some of the difficulties of dealing with the exact curvature can be eliminated by using a simpler matrix, at the cost of a larger cumulant correction (see Sec.~\ref{sec:corrections_calc}).

\section{Posterior evaluation} \label{sec:evaluation}
This section describes the implementation of the different terms in the posterior, Eq.~\ref{eq:posterior}. Two large log-determinants must be evaluated. The determinant of the CMB data covariance matrix is discussed in \ref{sec:covdet}, and is in fact fairly harmless. The second, the determinant of the lensing curvature matrix, discussed in \ref{sec:covh}, is more complex, and is the most expensive step of the entire implementation. However, we find that its dependence on the data realization is very weak, and can be neglected. For this reason, it only needs to be calculated once for a simulation suite. All terms must be evaluated at the best lensing map $\phi^*$. The production of this map together with the Wiener-filtered CMB maps $X^{\rm WF}$ is described in Sec.~\ref{sec:iteration}. Finally we give in Sec.~\ref{sec:corrections_calc} details on our generation of Gaussian lensing map samples with large and dense covariance matrices, as required for evaluation of the cumulant corrections.

\subsection{Data covariance determinant}\label{sec:covdet}

\begin{figure}
\centering
\includegraphics[width=0.5\textwidth]{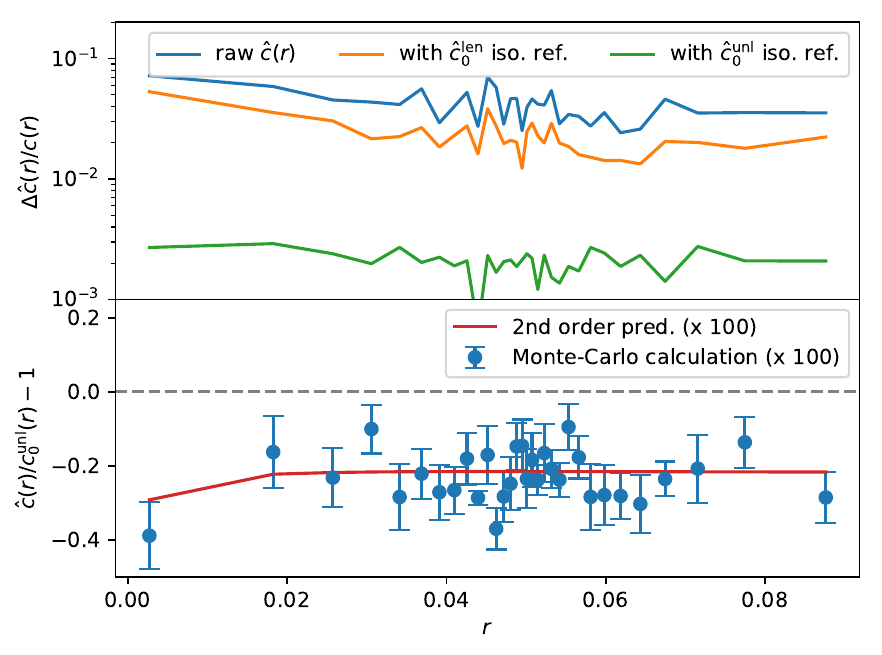}
 \caption{\label{fig:sncr}
 Upper panel: The three curves show the Monte-Carlo noise  $\Delta \hat c(r) /c(r)$ (one simulation equivalent) root variance for three estimators $\hat c(r)$ of the lensed data covariance $r$-derivative, Eq.~\eqref{eq:cracc}. The estimator accelerated by the isotropic reference estimator using unlensed spectra and neglecting lensing (green), is already accurate to well below a percent with a single Monte-Carlo estimate. Lower panel: The red curve shows the relative deviation of the analytic perturbative prediction to the isotropic, unlensed approximation $c^{\rm unl}_0(r)$ to $\hat c(r)$ (multiplied by a factor 100), Eq.~\eqref{eq:counl}. Blue points show estimates of the exact $c(r)$ from independent Monte-Carlo simulations at each $r$ point, where the iterative lensing solution $\phi^*(r)$ was obtained using 10 iterations starting from the quadratic estimator. it is apparent that the exact result is both very close to the isotropic estimation, and correctly captured by the analytical perturbative expansion. The error bars, independent from point to point, are the empirical standard deviations across 9 Monte-Carlo simulations. The figure was built using a data realization with input $r^{\rm in} = 0.05$ where the $r$-grid is densest.}
\end{figure}

How to estimate log-determinants of very large, dense matrices like $\rm{Cov}_\phi$? There is no trivial diagonalization, so that the large size of the matrix renders brute force methods totally useless. Possibilities include integral representations combined with Monte-Carlo evaluation of the diagonal \cite{Dorn:2015vfa}. We are only interested in the $r$-dependence. This allow us to simplify the problem quite a bit. We first note that
\beq
\nonumber
c(r) \equiv -\frac{\partial}{ \partial r }\frac 12 \ln \det \textrm{Cov}_{\phi^*(r)} = \frac 12 \Tr \:\textrm{Cov}^{-1}_{\phi^*(r)}  \partial_r\textrm{Cov}_{\phi^*(r)}
\enq
where $\partial_r \Cov$ only depends on the covariance matrix constructed from tensor CMB spectra\footnote{The derivative $\partial_r$ does not act on $\phi^*(r)$; we are interested in the explicit $r$-dependence only, for the reasons explained at the end of Sec.~\ref{sec:posterior}}. Using Monte-Carlo simulations, we can write an unbiased estimator as
\beq \label{eq:cr}
\hat c(r) = \frac 12\:X^\dagger \textrm{Cov}^{-1}_{\phi^*(r)}  \partial_r\textrm{Cov}_{\phi^*(r)}  \: X
\enq
where $X$ are unit spectra random variables with $\av{X_i X_j^\dagger} = \delta_{ij}$. It is easily seen that the estimator is unbiased. Its Monte-Carlo noise variance (when using Gaussian $X$) is simply the Fisher information on $r$, divided by the number of simulations used. We can further reduce the variance with the help of a reference ideal covariance matrix, denoted with a subscript $0$, for which we know the determinant: consider the modified estimator
\beq
\label{eq:cracc}
\hat c(r) - \hat c_0(r) + c_0(r) ,
\enq
the last term $c_0(r) \equiv \av{\hat c_0(r)}$ being known analytically. With a good isotropic approximation to the covariance we might expect to be able to reduce the MC noise of the original estimate substantially. Fig.~\ref{fig:sncr} shows in the upper panel the Monte-Carlo error $\Delta\hat c(r)/|c(r)|$ as a function of $r$, using the raw estimator $\hat c(r)$ (blue line) of Eq.~\eqref{eq:cr}, with the lensed spectra as reference (orange) in Eq.~\ref{eq:cracc}, and using unlensed spectra (green) as reference. More specifically, neglecting again the tensor $E$ contribution, these last two choices correspond to
\beq
c^{\rm len}_0(r) = \frac 12 \sum_\ell \frac{\left( 2\ell + 1\right) C_\ell^{BB, \rm tens.}}{rC_\ell^{BB, \rm tens.}+ C_\ell^{BB, \rm len} + C_\ell^{BB,\rm noise}}
\enq
in the lensed case, and
\beq \label{eq:counl}
c^{\rm unl}_0(r) = \frac 12 \sum_\ell \frac{\left( 2\ell + 1\right) C_\ell^{BB, \rm tens.}}{rC_\ell^{BB, \rm tens.} + C_\ell^{BB,\rm noise}}
\enq
in the unlensed case. In these two equations, $2 \ell + 1$ is short-hand notation for the exact number of modes of our flat-sky patch, and $C_\ell^{BB,\rm noise}$ is the beam-deconvolved noise spectrum. In the unlensed case, it is apparent that with a single simulation we can reach well below percent accuracy on the log-determinant.

In fact, for our experimental configuration at least, the isotropic, unlensed spectra determinant approximation is extremely accurate, and the small deviation from it can be very well captured by the leading term of a perturbative expansion in powers of $\phi$. This is shown in the lower panel of Fig.~\ref{fig:sncr}, which displays the relative deviation of the exact determinant $c(r)$ from its isotropic counterpart calculated with unlensed spectra (blue points with error bars) and the perturbative prediction (red line). More explicitly, the perturbative prediction is
\beq \label{eq:response}
\frac 12 \ln \det \textrm{Cov}_{\phi} = \frac 12 \ln \det \textrm{Cov}_{\phi = 0} + \frac 12\sum_{\boldsymbol{L}}R_{L} |\phi_{\boldsymbol{L}}|^2,
\enq
where the linear, mean-field response matrix $R$ is obtained in the appendix, and the first term on the right-hand side is evaluated with the unlensed CMB spectra. 
At first sight, it might appear counter-intuitive that the unlensed spectra prediction is so accurate, but this is easily explained. In the limit of vanishing instrumental noise and perfect resolution, the data covariance Eq.~\eqref{eq:cov} reduces to
\beq \nonumber
\Cov  \xrightarrow[\textrm{high res.}]{\textrm{low noise}}D_\phi C^{\rm{unl, fid}}D_\phi^\dagger.
\enq
Averaged over lensing maps this gives the CMB spectra \textit{lensed} by $C^{\phi\phi}$. However, taking the determinant one expects the factorization (for example after imposing a sufficiently high band-limit making all these operators square matrices)
\beq \nonumber
\ln \det \Cov \xrightarrow[\textrm{high res.}]{\textrm{low noise}} \ln \det D_\phi D_\phi^\dagger+ \ln \det C^{\rm unl, fid}.
\enq
The first determinant has no dependence on $r$ and is an irrelevant constant. Hence, usage of the \textit{unlensed} spectra gives under these conditions the exact result irrespective of the amount of lensing in the data. Fig.~\ref{fig:sncr} demonstrates that in our instrumental configuration, the mean-field response $R$ in Eq.~\eqref{eq:response} captures very well the tiny residual coupling between these terms, the transfer function and the noise matrix.

\subsection{Hessian determinant}\label{sec:covh}
The Hessian $H$ is defined as the second variation of the log-posterior $S$ of the lensing map, Eq.~\eqref{eq:Sexp}. Operationally speaking there are three terms $H \equiv H^{\rm dat} +  H^{\rm Cov} + H^{\rm Pri}$. The first originates from the quadratic piece of the CMB likelihood and is data realization dependent,
\beq
H^{\rm dat}_{\bL \bL'} \equiv \frac12 \left.\frac{\delta^2 }{\delta \phi_{\bL} \delta \bar \phi_{\bL'}}\Stdat \Covi \Stdat\right|_{\phi = \phi^*(r)}
\enq
the second also comes from the likelihood but is data-independent,
\beq
H^{\rm Cov}_{\bL \bL'} \equiv \frac12 \left.\frac{\delta^2 }{\delta \phi_{\bL} \delta \bar \phi_{\bL'}}\ln \det \Cov\right|_{\phi = \phi^*(r)},
\enq
and the third is the lensing map prior curvature, dominant on small scales where the data does not constrain the lensing deflection field. Under our choice of Gaussian statistics for $\phi$, this prior term is trivially given by
\beq
H^{\rm Pri}_{\bL \bL'} =\frac{\delta_{\bL \bL'}}{C_L^{ \phi\phi, \rm fid}}.
\enq
The calculation of the total Hessian determinant is the main difficulty to get the $r$-posterior. It plays a key role in shaping the final posterior and cannot be neglected, but is difficult to obtain exactly. However, as shown further below, the dependence of log-determinant on the data realization can be neglected for all practical purposes. Hence, this calculation only needs to be performed once for each model.

In general, we solve for $\ln \det H$ by coupling the same trace probing method described in Sec.~\ref{sec:covdet} to a double layered conjugate gradient inversion. We use for $H^{\rm Cov}$ our accurate approximation from Sec.~\ref{sec:covdet} for the covariance determinant, with the trivial result 
\beq
 H^{\rm Cov}_{\bL \bL'} = \delta_{\bL \bL'}R_L.
 \enq
The main operational difficulty lies in the application of $H^{\rm dat}$ to an arbitrary lensing potential map $\phi$. The details of this calculation are deferred to the appendix. There, it is shown that one can apply the Hessian matrix to a lensing potential vector at the cost of Wiener-filtering one pair of Stokes $Q, U$ maps. This operation is itself performed via conjugate gradient inversion. Using a simple diagonal preconditioner for the outer conjugate gradient inversion, we found that we need slightly above 10 iterations and thus the inverse filtering of the same number of polarization maps to conservatively solve for $\phi^\dagger H^{-1}\partial_r H \phi$ to five significant figures. The same operations must then be repeated for each Monte-Carlo $\phi$ used to probe the trace, and at each point $r$ of interest. For the area of $\sim 645$ square degrees and the resolution of $1.5$ arcminutes dealt with in this paper, a single probe of the trace at some $r$ point takes a couple of minutes. The final Monte-Carlo error on the PDFs depends on the $r$-density of trace-probing MC's, rather than the number of MC's per $r$ point. Hence it can be advantageous to use a dense $r$-grid where each point is sparsely trace-probed. For the PDFs shown in this work, where we reconstruct accurately the entire PDF shape, we typically use 64 $r$ points, denser near the peak, and $4$ MC's per point, which gives a crude but still reasonable estimate of the local error.

\begin{figure}
\centering
\includegraphics[width = 0.5\textwidth]{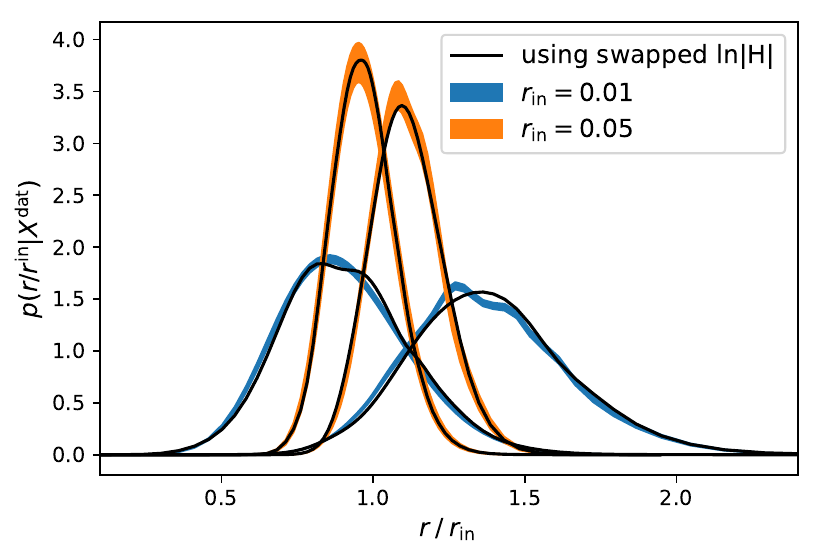}
 \caption{\label{fig:lndetH}\textit{Data dependence of the Hessian log-determinant.} Comparison of the $r$-posterior reconstruction on two simulated data maps spanning $645 \deg^2$ with  $2.1\mu $K-arcmin. polarization sensitivity. The filled colors use the expensive, realization-dependent curvature determinant estimate. The width shows the $2\sigma$ Monte-Carlo uncertainties from the determinant derivative calculation. For each realization, errors on opposite sides of the peak of the posterior are strongly anti-correlated. The black lines show the same reconstructions but swapping the log-determinant between the two realizations (black lines), with virtually identical results. For the cases we investigated, the determinant needs only be calculated on a single data realization.}
\end{figure}

Figure~\ref{fig:lndetH} shows a couple of posterior reconstructions on simulated data. One pair with input $r^{\rm in} = 0.01$ (orange), and one pair with $r^{\rm in} = 0.05$ (blue). The filled colours shows the reconstructions using the realization-dependent Hessian estimate, where the width displays the $2\sigma$ uncertainties resulting from the Monte-Carlo determinant evaluation. These errors are independent only for the estimation of the posterior $r$-derivative: after normalization to build the posterior itself, this produces a strong anti-correlation of errors between points on opposite side of the peak on this figure. The black lines show the corresponding curves, but with Hessian estimation swapped between the two realizations. Summary statistics are virtually identical, and this demonstrates that for all practical purposes, we can neglect the data dependence of the log-determinant. However, the dependence on the lensing potential $\phi^*$ remains, and we do not offer in this work a trivial, fully isotropic approximation.

\begin{figure}
\centering
\includegraphics[width = 0.5\textwidth]{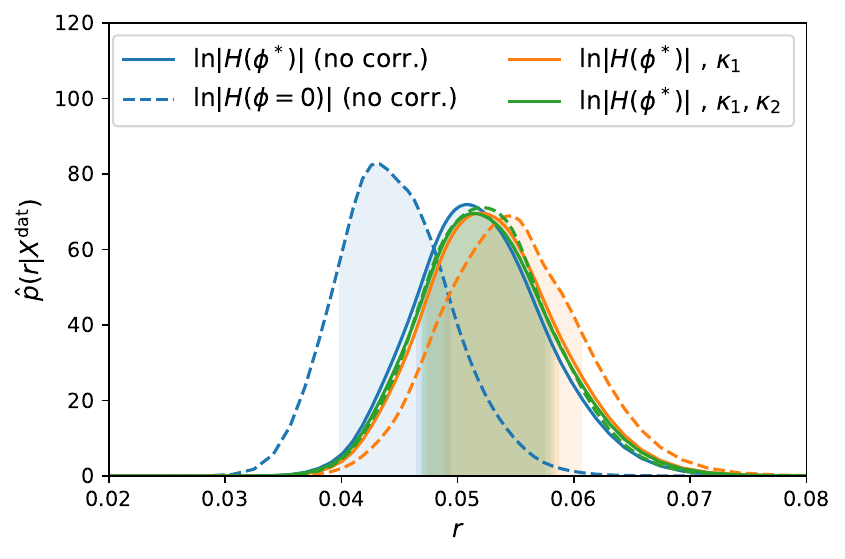}
 \caption{\label{fig:corrections}\textit{Corrections to posterior and alternative expansion}. Tensor-to-scalar ratio $r$ posterior estimates for one data realization (with input value $r_{\rm in} = 0.05$), without any cumulant correction (solid blue), including first (solid orange) and second (solid green) cumulant corrections. The dashed lines show the same curves, but using a different posterior expansion scheme, with a less accurate but simpler curvature matrix neglecting the lensing. The agreement between the two methods is very good and provides a good consistency check, but in the latter case the inclusion of the cumulant corrections becomes mandatory. The dashed bands contain 68\% of the probability.}
\end{figure}
\subsection{Iterative lens estimation}\label{sec:iteration}
The first step in the $r$-posterior reconstruction is always the calculation of the iterative solution $\phi^*$ together with the Wiener-filtered CMB maps $X^{\rm WF}$. We follow Ref.~\cite{Carron:2017mqf} very closely, maximizing the lensing map posterior probability for the lensing deflections using a quasi-Newton iterative descent. At each iteration step, the current lensing estimate induces a mean-field ($\av{\phi}$, in standard lensing estimation terminology) term to subtract from the quadratic estimate, originating from the first variation (with respect to $\phi$) of the covariance matrix determinant. This term is small on all scales for lens reconstruction from polarization, and we use the same accurate analytical approximation discussed in Sec.~\ref{sec:covdet} and in the appendix instead of simulations.
\subsection{Higher-order cumulants calculation}\label{sec:corrections_calc}
Evaluation of the correction terms in Eq.~\eqref{eq:cumulants} requires sampling Gaussian lensing maps with inverse covariance given by the curvature matrix chosen for the posterior expansion. This matrix is very large and has no trivial structure, hence is difficult to sample from, and standard generic methods such as Cholesky decomposition are useless. It is possible, however, to apply this matrix to a lensing map in reasonable time, and powerful methods can be used to construct samples with the correct covariance structure using this property. We proceed by deforming continuously the inverse square root of H as follows\citep{Allen:2000:NAP}. We connect $H$ to the identity matrix $\mathbb I$ defining 
\beq
H_t \equiv \mathbb I + t(H - \mathbb I),
\enq
and introduce $\phi_t \equiv H^{-1/2}_t \phi_0 $ where $\phi_0$ is a unit variance Gaussian vector of the appropriate dimensionality. By definition, $H_{t=1} = H$ and $\phi_{t = 1}$ has covariance $H^{-1}$. Differentiating gives the following ordinary differential equation (ODE) for $\phi_t$:
\beq\label{eq:ODE}
\frac{d\phi_t}{dt} = -\frac 12 \left(H - \mathbb{I}\right) H_t^{-1}\phi_t
\enq
The solution of this ODE at $t= 1$ is then by construction the desired Gaussian sample\footnote{After discretization, the algorithm is reminiscent of the Van Cittert deconvolution of image analysis \cite{van1931einfluss, starck2002deconvolution}.}. We have used this machinery to test corrections to our posterior approximation Eq.~\eqref{eq:posterior} on one data realization. In this case,
\beq
H_{\bL\bL'} = \left.\frac{\delta^2 S[\phi]}{\delta \phi_{\bL} \delta \bar \phi_{\bL'}}\right|_{\phi = \phi^*(r)}.
\enq
As a consistency check we also checked the alternative choice of curvature where the lensing deflections are neglected,
\beq
H_{\bL\bL'} = \left.\frac{\delta^2 S[\phi]}{\delta \phi_{\bL} \delta \bar \phi_{\bL'}}\right|_{\phi \equiv 0}.
\enq
The second case considerably speeds-up the calculation of the curvature log-determinant, but the posterior is expected to be less accurate. Figure~\ref{fig:corrections} shows the different $r$-posterior estimates, for one data realization with input $r^{\rm in} = 0.05$. The solid lines shows our baseline approximation, without corrections (blue), with first cumulant (orange) and with first two cumulants (green). We have used 4 Monte-Carlo samples per $r$-point, on a grid devised as just described in Sec.~\ref{sec:covh}. We use in all cases a standard low order Bulirsch-Stoer algorithm \cite{Press1996} to solve the ODE defined in Eq.~\eqref{eq:ODE}. The bracketed inverse is solved with conjugate-gradient descent. The corrections have a completely negligible effect on the posterior. The dashed lines show the same curves with the choice of unlensed curvature. The agreement, including the corrections, is very good. However, the leading approximation is clearly worse and the corrections are essential in bringing the two expansion schemes in agreement. In the remainder of this paper we stick to the lensed curvature matrix, and safely neglect the cumulant corrections.
\section{Nominal and delensed $B$-mode band powers}\label{sec:bandpowers}
We now describe our reconstruction of $r$ from raw and delensed band powers. The likelihood of CMB data temperature and polarization band powers has been discussed at length in several places already, and efficient parameterizations both at low and high multipoles are well known \cite[e.g.][]{Efstathiou:2003dj, Hamimeche:2008ai}. Only preliminary work exists, however, for delensed band powers~\cite{Namikawa:2015tba}. After performing consistency checks described in the next section, we decided to use the Hamimeche and Lewis (HL) likelihood \cite{Hamimeche:2008ai}, using analytical band-power predictions and covariance matrices as described below. We note that other possibilities such as the parametrization of Ref.~\cite{Smith:2005ue} seem to work just as well. We consider $B$-mode multipoles only up to $\ell_{\rm max} = 200$, and use as baseline the maximal number of bins (74) that our flat-sky mode structure allows when building the $r$-posteriors.
 
 All of our $B$-spectra likelihoods may eventually be written
\beq
\nonumber
-2\ln p(r | \hat C^{BB}) \equiv g(x_\ell)\hat C^{BB}_{\ell}\:\Sigma^{-1}_{\ell{\ell'}}\:\hat C^{BB}_{\ell'} g(x_{\ell'})
\enq
with $g(x) \equiv \textrm{sign}(x-1)\sqrt{2(x - \ln x - 1)}$ and $x_\ell(r) = \hat C^{BB}_\ell / C^{BB}_\ell(r)$. These power spectra include tensor contribution, residual lensing power and beam-deconvolved noise. The predictions of the spectra $C^{BB}_\ell(r)$ are evaluated analytically, using the perturbative expression for the lensing $B$-mode power to linear order in the appropriate lensing spectrum. For the nominal band powers we use the fiducial lensing spectrum $C_L^{\phi\phi, \rm fid}$. In the case of the quadratic estimator or iteratively delensed band powers, we first empirically measure the cross-correlation coefficient squared $\rho^2_L$ of the reconstructed lensing maps to the true input from a small set of simulations. We then use as lensing spectrum $C_L^{\phi\phi, \rm fid}\left(1 - \rho_L^2\right)$. We neglect the tiny lensing effects on the tensor spectra. The lensing maps themselves are obtained as described in Sec.~\ref{sec:iteration}, with the difference that we exclude the tensor modes scales $\ell  \leq 200$ to build the tracers. Including these modes would dramatically complicate the analysis: if the delensed data and lensing tracer contain common multipoles, the delensed $B$-modes would contain very strong, spurious delensing signature that originate from disconnected 4-points and 6-points statistics of the CMB \cite{Teng:2011xc, Carron:2017vfg, Namikawa:2017iak}. The reason is that the lensing estimator cannot distinguish between true and random lensing signatures in the data (such as shear or magnification) and all of these signatures are removed after delensing, leading to a spuriously unlensed-looking CMB at the two-point level. The loss of signal to noise in the lensing tracer by excising the largest angular scales is small: the delensing efficiency $\rho_L^2$ is reduced by roughly 1\% on the most relevant scales $L \sim 500$.

To delens the CMB, we use a simple remapping technique. From the Wiener-filtered lensing deflection estimate $\boldsymbol{\alpha}$, we simply remap the Stokes parameter (after beam deconvolution, and discarding multipoles higher than 2000) according to the lensing inverse deflection field $\boldsymbol{\alpha}^{-1}$, defined through the condition that deflected points $\boldsymbol{x} + \boldsymbol{\alpha}(\boldsymbol{x})$ get remapped back to themselves
\beq
\bf x + \boldsymbol \alpha^{-1}(\bf x + \boldsymbol \alpha(\bf x)) \equiv \bf x.
\enq
This inversion is performed exactly following Ref. \cite{Carron:2017mqf}, using a fast-converging Newton-Raphson solver. This delensing technique differs operationally and conceptually from the template method used for instance by the SPTpol team on their polarization data \cite{Manzotti:2017net}, which uses a Wiener-filtered $E$-mode map to build a $B$ template, then subtracted from the data. The filters of the template method are optimized to minimize the resulting $B$-power at linear order \cite{Sherwin:2015baa}. While the remapping method naturally contains higher-order terms, the noise is also remapped, which can result in higher total $B$ power if this is high such as for Planck data  \cite{Carron:2017vfg, Aghanim:2018oex}. However, we found that there is basically no difference between the two methods for the low levels of noise here. By default we include the noise remapping into our delensed $B$-power predictions, but this is very small. Finally, there are no $E/B$ separation complications \cite{Lewis:2001hp, Smith:2006vq} since we use periodic patches throughout this paper and the spectra estimator $\hat C^{\rm BB}_\ell$ are trivial to build from the delensed or nominal Stokes maps.

For the covariance matrix, we use the simple approximation  \cite{Smith:2004up, BenoitLevy:2012va} combining the Gaussian part with corrections according to the lensing kernels:
\beq
\begin{split}
\Sigma_{\ell_1\ell_2} = &\:\delta_{\ell_1 \ell_2} \sigma^2_{\ell_1}(C^{BB}) \\ &+ \sum_{\ell}\frac{\partial C_{\ell_1}^{BB}}{\partial C^{EE, \rm unl}_\ell}\sigma^2_{\ell}(C^{EE, \rm unl})\frac{\partial C_{\ell_2}^{BB}}{\partial C^{EE, \rm unl}_\ell} \\
&+ \sum_{L}\frac{\partial C_{\ell_1}^{BB}}{\partial C^{\phi\phi}_L}\sigma^2_{L}(C^{\phi\phi})\frac{\partial C_{\ell_2}^{BB}}{\partial C^{\phi\phi}_L}
\end{split}
\enq
with Gaussian spectrum variance
\beq
\sigma^2_\ell(C^{XX}) \equiv \frac{2 \left(C_\ell^{XX}\right)^2}{2\ell + 1}
\enq
where $2\ell + 1$ really stands for the number of multipoles in our flat-sky patch. The derivatives are evaluated to first order in $C_L^{\phi\phi, \rm fid}$, or in $C_L^{\phi\phi, \rm fid}(1 - \rho_L^2)$ for the delensed band powers. On the relevant scales $\ell \le 200$, the covariance correlation coefficients are small and of minor importance, and the perturbative expansion is accurate enough. Collecting reconstructions from simulations, we checked that the diagonal of the covariance matrices matches the prediction to within a couple of percent at least. 


\section{Results} \label{sec:results}
\begin{figure}[h!]
\centering

\includegraphics[width = 0.45\textwidth]{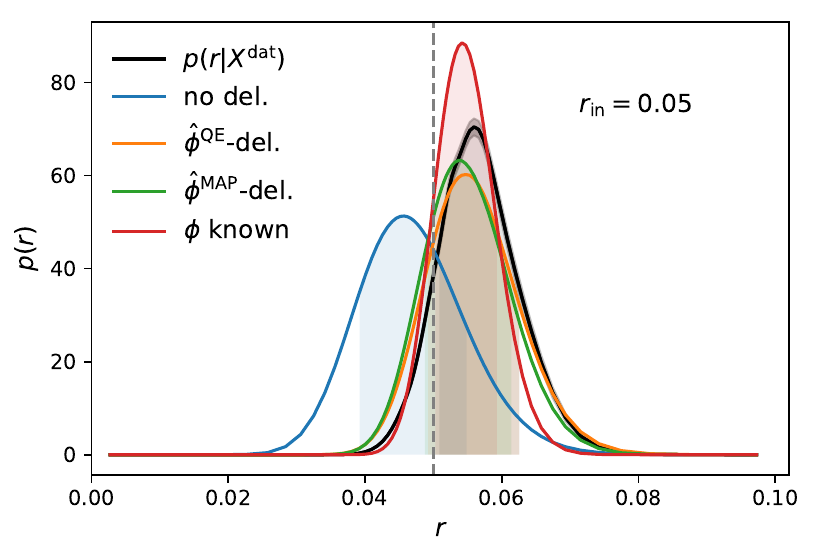}
\includegraphics[width = 0.45\textwidth]{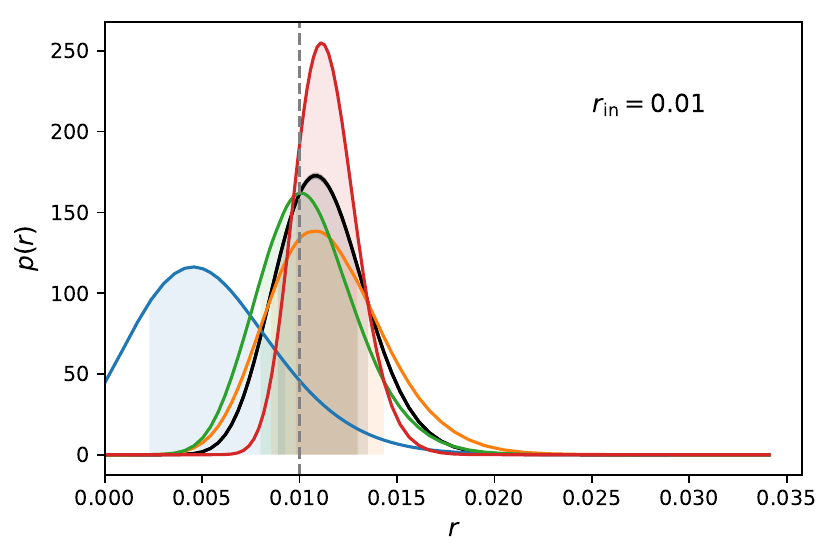}
\includegraphics[width = 0.45\textwidth]{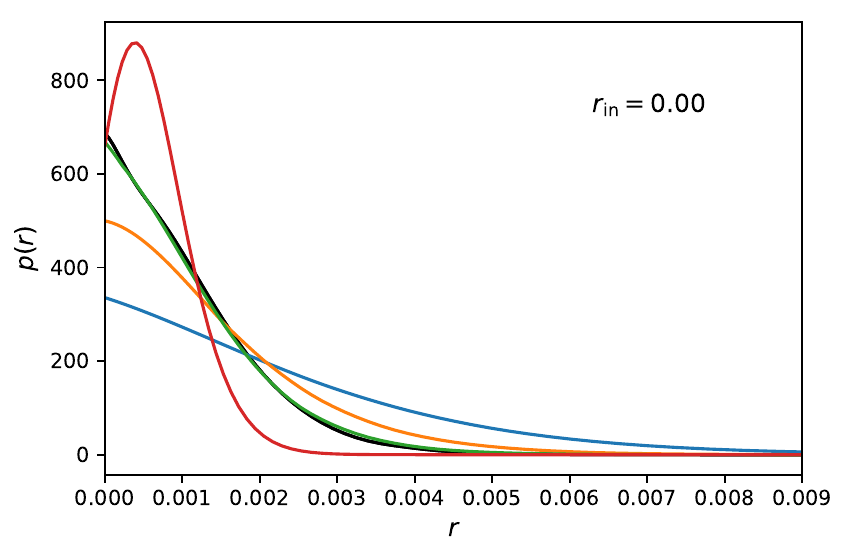}
 \caption{\label{fig:pdfplot}Posterior constraints on the tensor-to-scalar ratio $r$ for different reconstruction methods on one data realization with an input $r_{\rm in} = 0.05, 0.01$ and $0$ from top to bottom. The black line shows our new estimator. The green, orange and blue lines show the constraints from nominal, quadratic estimator delensed and iteratively delensed $B$-mode band powers respectively, with likelihood built as described in the text. For comparison, the red line shows the case of perfect knowledge of the input lensing map, with posterior given by Eq.~\eqref{eq:phiknown}. In the upper two panels, the dashed bands contain 68\% of the probability.}
\end{figure}

\begin{table}\caption{\label{table:stats}Summary statistics comparison on tensor-to-scalar ratio constraints averaged over our data realizations, for the different methods tested in this work and three input values of tensor modes as indicated in the first row. The reconstructions are performed on maps of $645 \deg^2$ with $2.12 \mu$K-arcmin. polarization noise. The first three rows show the results of $B$-mode band powers likelihood analysis, without delensing, with quadratic estimator delensing and with iterative delensing. The next row shows the constraints from the exact posterior density function on $r$, obtained as described in this work. The last row shows the case of a perfectly known lensing map for comparison. We quote twice the standard deviation for the first two columns where $r$ is well constrained, and the $95\%$ confidence limit in the last column.}
\begin{tabular}{|l|lll|}
\hline
\hline
100\:$r^{\rm in}$ & 5.0  & 1.0 & 0.0 \\
\hline
band powers, no delensing  & 1.65 ($2\sigma$) & 0.83 ($2\sigma$) & 0.75 (95\% c.f.)\\
band powers, $\phi^{\rm QE}$-delensing & 1.27  & 0.56  & 0.40 \\
band powers, $\phi^{\rm MAP}$-delensing & 1.18  & 0.49  & 0.31 \\
Exact posterior & 1.14 & 0.46 & 0.29\\
\hline 
\hline 
$\phi$-known posterior & 0.87 & 0.30 & 0.12\\

 \hline
\end{tabular}
\end{table}

\begin{figure*}
\centering
\includegraphics[width = \textwidth]{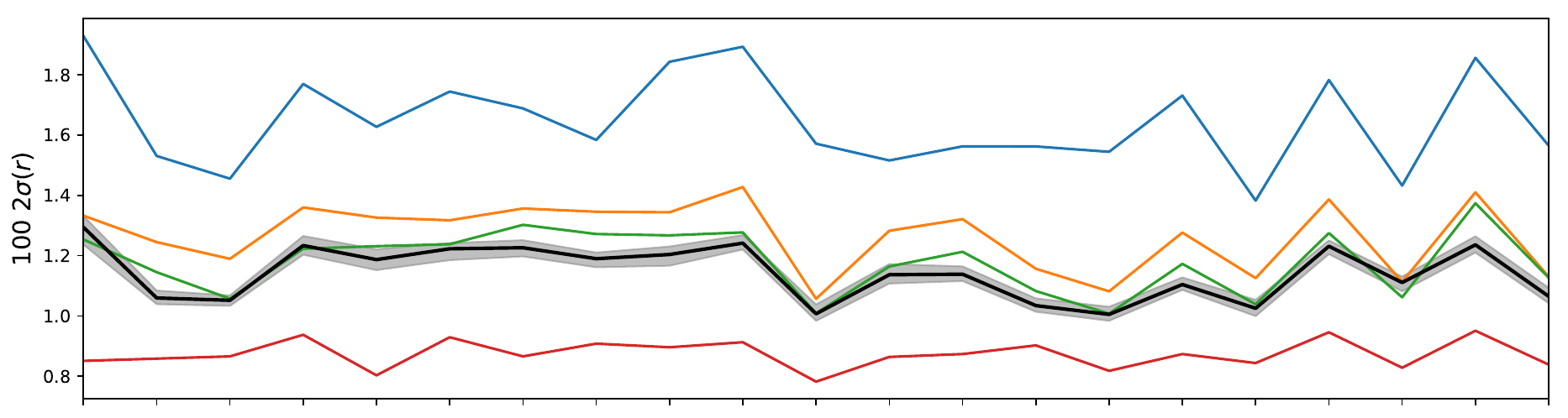}
\includegraphics[width = \textwidth]{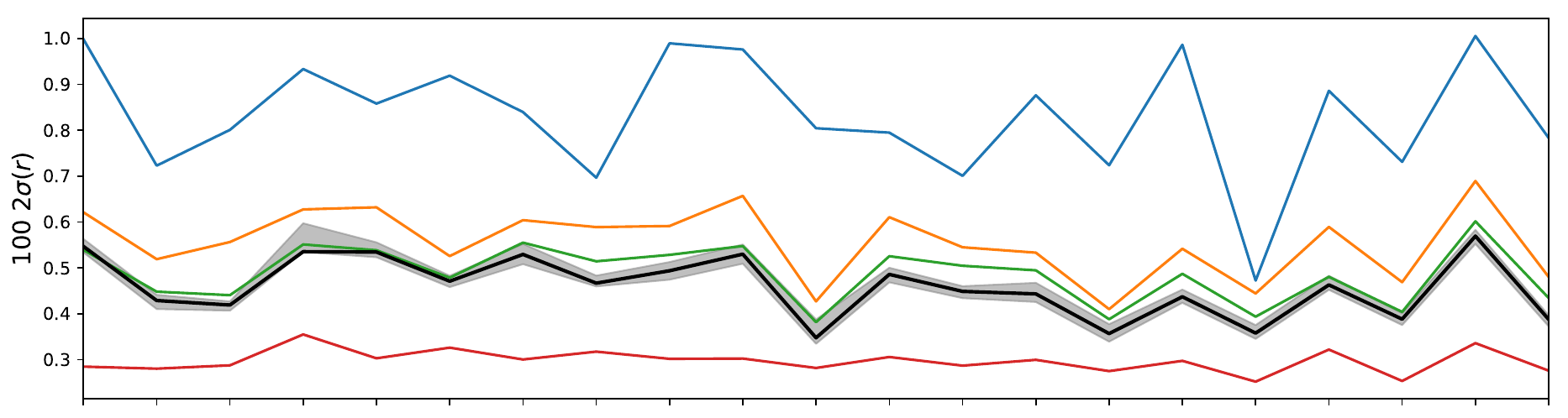}
\includegraphics[width = \textwidth]{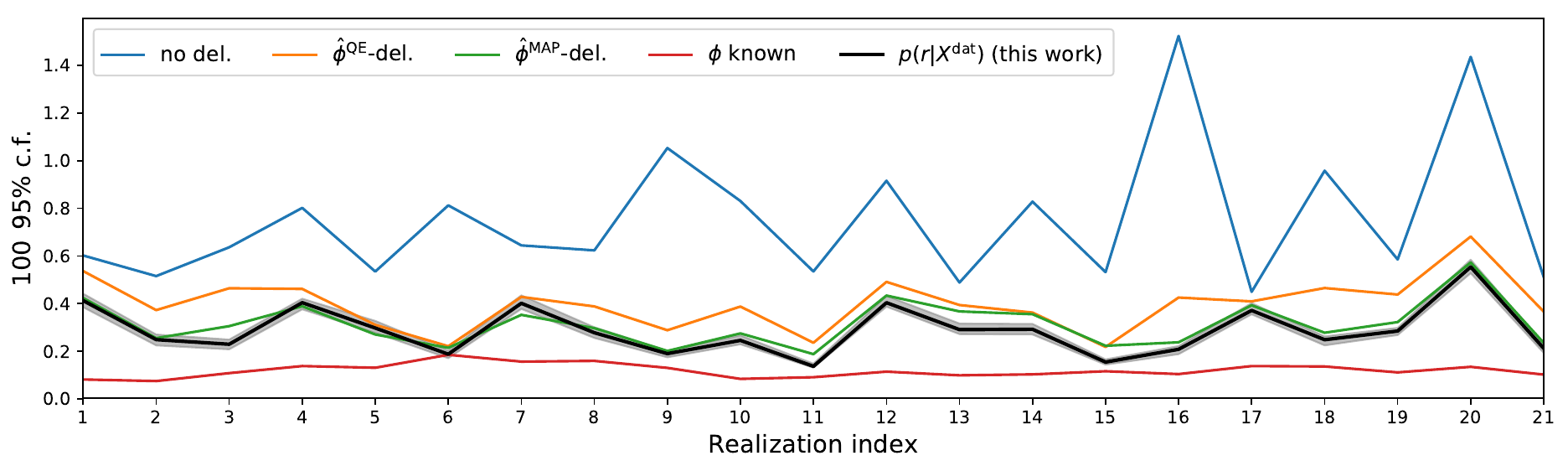}
\caption{\label{fig:pdfbandplot} Comparison of constraints on the tensor-to-scalar ratio $r$ obtained from band-power analysis and the $r$ posterior obtained in this work (black), for 21 simulated data maps. Shown are constraints from nominal $B$-mode band powers (blue), quadratic estimator delensed band powers (orange), and iteratively delensed band powers (green). The red curves show the constraints when the lensing deflection potential is perfectly known for comparison. The grey shaded area shows the $2 \sigma$ uncertainty on the constraint due to the Monte-Carlo evaluation of one of the deflection Hessian curvature determinant in the posterior. The three panels use  maps with different input tensor amplitudes $r_{\rm in} = 0.05, 0.01$ and $0$ (top to bottom). We quote 2 standard deviations (multiplied by 100) in the first two panels, where the tensor modes are well constrained, and the 95\% confidence limit (also multiplied by 100) in the lowest panel with vanishing input tensor amplitude. All reconstructions are performed on $645 \deg^2$, with a white polarization noise level of $2.12 \mu $K-amin.}
\end{figure*}
 We have performed our baseline posterior reconstruction, including realization-dependent determinant calculation, on 21 CMB simulations for each of the 3 different levels $r_{\rm in} = 0.05, 0.01, 0$ of tensor modes advertised in the introduction. For each realization, we also performed likelihood analyses from the quadratic-estimator delensed, iteratively delensed $B$-mode band powers and with no delensing. Fig.~\ref{fig:pdfbandplot} shows our main results. For each simulation the constraints obtained from either our posterior $p(r|\Stdat)$ approximation are displayed (black), together with those obtained from the different band-power analyses (coloured lines). Blue shows the results from the nominal (no delensing) band-power likelihood, orange after delensing with the Wiener-filtered quadratic estimator $(\phi^{\rm QE})$, and green after delensing with the iterative (formally, the maximum a posteriori (MAP) solution $\phi^{\rm MAP}$) lensing solution . Finally, for comparison the red line shows the constraints achievable if the lensing deflections were known perfectly. Up to irrelevant constants, this latter case is given by
 \beq \label{eq:phiknown}
\ln p(r, |\Stdat,\phi^{\rm in}) \propto -\frac 12 \Stdatt \rm{Cov}_{\phi^{\rm in}}^{-1} \Stdat - \frac 12 \ln \det \rm{Cov}_{\phi^{\rm in}},
 \enq
 where $\phi^{\rm in}$ is the lensing potential input to data realization $\Stdat$. The three panels shows $r_{\rm in} = 0.05, 0.01$ and $0$ from top to bottom. In the first two panels the quoted numbers are the $2\sigma $ width of the PDF, and the lowest panel shows the 95\% confidence upper limit.  The grey area around the black line shows the $2\sigma $ uncertainty originating from the Monte-Carlo measurement of the Hessian determinant. We evaluate these errors simply by sampling an ensemble of posteriors from the Monte-Carlo log-determinant's errors. Fig.~\ref{fig:pdfplot} shows explicitly the $r$ posterior PDF for one of the realizations using the same colour scheme.  As visible there, all reconstructions show skewness and other non-Gaussian features, hence these summary statistics are only an imperfect representation of the constraints on $r$.
 
 All of our exact posterior estimates mean value lie within 1.8$\sigma$ ($r^{\rm in} = 0.05$) of the true value. The series of $r^{\rm in} = 0.01$ simulations shows a single outlier, with recovered posterior mean $100\:\av{r} = 0.53$ formally lower than the input by $2.5\sigma$, but with a pdf showing substantial skew with no strong evidence of an anomaly. We can try and isolate a possible systematic bias in our posterior approximation by combining all posteriors as if they were independent measurements of $r$. Doing so, we obtain the constraints
\beq
100 \:\hat r = 5.15 \pm 0.12 \quad(100 \:r^{\rm in} = 5, \rm combined)
\enq
\beq
100 \:\hat r = 1.04\pm 0.05 \quad (100 \:r^{\rm in} = 1, \rm combined),
\enq
both perfectly consistent with the input values. This constrains a systematic bias to a small fraction of a standard deviation. The same test on all the band-power likelihoods or the known-$\phi$ likelihood (Eq.~\ref{eq:phiknown}) also shows consistency.

It is apparent that there is significant spread in the summary statistics displayed on Fig.~\ref{fig:pdfbandplot}. As expected, this is more pronounced for the nominal band-power analysis which has the weakest constraints. While the different methods constraining power do rank as expected on average, the results using the iteratively delensed band powers in green and our full posterior reconstruction in black are very close, with a few reconstructions showing nominally slightly worse summary statistics from the posterior reconstruction. These cases seem compatible with the errors on the posterior. Also, since the exact band-power likelihood is intractable, it is very difficult to assess precisely the impact of our band power-likelihood approximation on each realization. In all cases, iterative delensing does perform better than quadratic estimator delensing, but the improvement shows substantial realization dependence as well, and in a couple of cases can be absolutely minimal.
Table~\ref{table:stats} shows the summary statistics averaged across all these simulations. We find that the exact posterior outperforms the iteratively delensed band powers by 3\%, 6\% and 8\% for $r^{\rm in} = 0.05, 0.01$ and $0$ respectively.

\section{Summary}\label{sec:summary}
Unless the primordial $B$-mode power produced during inflation is very large, sophisticated analysis techniques such as delensing will be essential to provide best constraints on primordial gravitational waves. We have presented a new estimator, based on a close approximation to the exact posterior of the tensor-mode amplitude. By careful Monte-Carlo investigations of corrections to the approximation, we have demonstrated that it is unbiased and very close to optimal, providing the tightest possible constraints on primordial gravitational waves from CMB data. The estimator uses fast, joint estimation of the best lensing deflection map and of the unlensed CMB.

This first investigation used a simplified setting, including periodic sky patches, no analysis mask, and usage of homogeneous noise, facilitating both the iterative lens reconstruction and the unlensed $E$-$B$ recovery from the observed lensed Stoke polarization data. These assumptions did not play a key role in obtaining our results, since the presence of the lensing deflections and data realization dependence break isotropy and prevent the existence of trivial basis to work with. Usage of Monte-Carlo simulations and inversion methods akin to conjugate-gradient seem unavoidable. Lens reconstruction and Wiener-filtering on masked data have already been demonstrated successfully~\cite{Carron:2017mqf} with the same methods, at the cost of a manageable increase in execution time.

The posterior reconstruction has dependency on two aspects of the cosmological model used to define the CMB likelihood: the unlensed CMB scalar perturbations spectra and the lensing potential power spectrum. Changes in the lensing spectrum (or lensing deflection map prior) impacts slightly the optimal lens reconstruction, and changes in the scalar spectra the unlensed $E$ and $B$ polarization field reconstruction. Hence, formally, usage of a slightly different fiducial model might lead to a slightly different result. However, all these power spectra are extremely well constrained in practice from observations, including the lensing spectrum, so this is unlikely to bring significant biases. Furthermore, if necessary it is possible to include the uncertainty in the power spectra in the posterior, by obtaining the linear response to the spectra and extending in this way the posterior density, in analogy to the way state-of-art lensing spectrum reconstruction likelihoods are built~\cite{Ade:2015zua, Sherwin:2016tyf}. One can also go a step further and marginalize directly over these using the empirical spectra, as demonstrated by Ref.~\cite{Aghanim:2018oex}. We also restricted our analysis to the extraction of the tensor-to-scalar ratio assuming a fixed template shape of the tensor spectrum. However, this is not a limitation of this approach; using high-quality observations the obvious generalization of this framework will be able to distinguish features as well.

We have compared the performance of our estimator to that of more traditionally planned $B$-mode band powers extraction. We found standard, well-demonstrated analytical likelihood models are able to describe meaningfully the delensed band powers, and we have used these likelihoods on nominal, quadratic estimator delensed and iteratively delensed band powers. The performances of these delensed band powers do match naive expectations. For the configuration studied here, targeting the recombination peak of the $B$-mode spectrum with noise levels in line with expectations from a CMB stage-IV experiment, our new estimator does outperform the iteratively delensed band powers by a realization-dependent amount, also depending on the exact value of $r$, reaching $8\%$ on average for small values. Producing the posterior PDF for $r$ as we did is more expensive numerically than producing band powers. Nevertheless, we demonstrated in this paper that the analysis was possible, providing constraints optimal by construction, and improving prospects on detecting a tantalizing component of modern cosmology.
\begin{acknowledgements}
I thank Antony Lewis for many discussions and useful comments on the manuscript, Anthony Challinor for early discussions that eventually lead to the material presented this paper, and Jes\'us Torrado for discussions on Gaussian vectors with large dense covariance matrices. The research leading to these results has received funding from the European Research Council under the European Union's Seventh Framework Programme (FP/2007-2013) / ERC Grant Agreement No. [616170]. This research used resources of the National Energy Research Scientific Computing Center, a DOE Office of Science User Facility supported by the Office of Science of the U.S. Department of Energy under Contract No. DE-AC02-05CH11231
\end{acknowledgements}
\bibliography{lensingbib}

\appendix
\onecolumngrid


\newcommand{\Res}[0]{\bar X}
\newcommand{\Xdag}[0]{X^{\rm{WF},\dagger}}
\newcommand{\X}[0]{X^{\rm{WF}}}
\newcommand{\vecell}[0]{ {\boldsymbol{\ell}} }

\section{Lensing map posterior curvature}\label{app:curvature}

This appendix describes in more details the lensing map posterior curvature matrix $H = H_r[\phi^*(r)]$, defined as the matrix of second variation of the lensing map log-posterior,
\beq\label{eq:curvature}
H_{\bL\bL'} \equiv \left.\frac{\delta^2 S[\phi]}{\delta \phi_{\bL} \delta \bar \phi_{\bL'}}\right|_{\phi = \phi^*(r)}, \textrm{   with } S[\phi] \equiv \frac 12 \Stdat \Covi \Stdat + \frac 12 \ln \det \Cov + \frac 12 \sum_{\boldsymbol{L}} \frac{\left|\phi_{\boldsymbol{L}}\right|^2}{C^{\phi\phi, \rm fid}_L}.
\enq
where the first two terms on the right-hand side come from the lensing map likelihood, and the last term from the Gaussian prior. In particular, in this appendix we obtain the result that the data-dependent part of the matrix can be applied in linear time to a vector (at the cost of one CMB Wiener-filtering). This makes the calculation of its log-determinant and inverse possible. The data-independent part poses no particular problems. The Gaussian prior curvature is trivial, and for the second term in Eq.~\eqref{eq:curvature} we adopt the same very accurate approximation as in Sec.~\ref{sec:covdet},
\beq\label{eq:mfresp}
 \frac 12 \ln \det \Cov \simeq  \frac 1 2 \ln \det \textrm{Cov}_{\phi \equiv 0} + \frac 12 \sum_{\bL \bL'}\phi_\bL \bar \phi_\bL'\left. \frac{\delta^2\frac 12 \ln \det \Cov}{\delta \phi_\bL\delta \bar \phi_{\bL'} }\right|_{\phi \equiv 0}  \equiv \frac 1 2 \ln \det \textrm{Cov}_{\phi \equiv 0} + \frac 12 \sum_{\boldsymbol{L}} R_{L} \left|\phi_{\boldsymbol{L}}\right|^2,
 \enq
where $R_L$ is the $\phi$-induced mean-field linear response (hence also the desired curvature term), obtained analytically following Ref.~\cite{Carron:2017mqf} and reproduced briefly at the end of this appendix. 

To obtain the data-dependent part of the curvature, we calculate for convenience the curvature $H_{\rm dat}^{ab}(\vx, \vy)$ with respect to the two position-space deflection components $\alpha_a(\vx),\alpha_b(\vy)$ of the flat sky, where $\alpha = \nabla \phi$. Once this is done, it is straightforward to apply the matrix to a lensing potential vector by expanding it into these two components, applying $H^{ab}$, and re-projecting eventually onto the potential Fourier harmonics. The data-dependent curvature splits into two terms which we describe next:
\beq \label{eq:Hdat}
\begin{split}
H_{\rm dat}^{ab}(\vx,\vy) &= \frac 12 \Stdat\cdot\Covi \frac{\delta\Cov}{\delta \alpha _a(\vx)} \Covi\frac{\delta\Cov}{\delta \alpha_b(\vy)}\Covi \Stdat + (a,\vx\leftrightarrow b,\vy) \quad \left( \equiv 2 H^{ab}_{F}(\vx, \vy) \right) \\
&- \frac 12 \Stdat\cdot \Covi\frac{\delta^2\Cov}{\delta \alpha_a(\vx)\delta\alpha_b(\vy)}\Covi\Stdat \quad \left( \equiv H^{ab}_{\rm det}(\vx, \vy) \right).
\end{split}
\enq
When the data covariance precisely matches  $\Cov$ (that is, for our simulations, for fiducial $r$ values close to the true value in $\Stdat$).
the term $H_F$ coincides, on average, with the Fisher information matrix on the displacement field. Under these conditions, and setting $\phi \equiv 0$, it is precisely the isotropic inverse lensing reconstruction noise $N^{(0)}_{\boldsymbol{L}}$ for the deflection angle calculated with the unlensed CMB spectra in the weights\cite{Okamoto:2003zw}. The second term $H_{\rm det}$ is subdominant: on average, and again when the data covariance matches $\Cov$, this term would be cancelled by a contribution from $R_{\boldsymbol{L}}$. This term $R_{\boldsymbol{L}}$ would also cancel one of the two $H_F$ factors in Eq.~\eqref{eq:Hdat}. The complete relation Eq.~\ref{eq:curvature} for the lensing potential Fourier modes under these conditions is
\beq
\av{H_{\boldsymbol L \boldsymbol L'}}_{\Stdat} = \delta_{\boldsymbol L \boldsymbol L'} \left(\frac{1}{N^{(0)}_{\boldsymbol L} } + \frac{1}{C^{\phi\phi,\rm fid}_{\boldsymbol L} } \right) \quad \left(\textrm{when } \phi \equiv 0,\textrm{  and  } \av{\Stdat X^{\rm dat, \dagger}} = \rm{Cov}_{\phi = 0} \right),
\enq
which can serve as a useful consistency check of the data-dependent curvature calculation.

To proceed, we need the first and second derivatives of the covariance. To simplify notation, we introduce
\beq\label{eq:cov1stvar}
\xi^\phi \equiv D_\phi C^{\rm unl} D_\phi^{\dagger},\quad \xi_{,a}^\phi \equiv D_\phi \nabla_a C^{\rm unl} D^{\dagger}_\phi,\quad \textrm{and}\quad  \xi_{,ab}^\phi \equiv D_\phi \nabla_a\nabla_b C^{\rm unl} D^{\dagger}_\phi
\enq
where the operator $\nabla_a C^{\rm unl}$ is defined by the block-diagonal matrix Fourier representation $\delta_{\bl \bl'} i \bl_a \begin{pmatrix} C^{EE, \rm unl}_\ell & 0 \\  0 & r\:C^{BB, \rm tens.}_\ell\end{pmatrix}$, and similarly for $\nabla_a \nabla_b C^{\rm unl}$ with an additional factor $i \bl_b$.
The following relations hold
\beq
\frac{\delta{\Cov}^{ij}}{\delta \alpha_a(\vx)} = {\mathcal{B}}(\vx_i,\vx)(\xi_{,a}^\phi {\mathcal{B}}^\dagger)(\vx,\vx_j) - ({\mathcal{B}}\xi^\phi_{,a})(\vx_i,\vx)({\mathcal{B}}^\dagger)(\vx,\vx_j),
\enq
and
\beq \label{eq:cov2ndvar}
\begin{split}
\frac{\delta^2{\Cov}^{ij}}{\delta \alpha_a(\vx)\delta \alpha_b(\vy)}& = {\mathcal{B}}(\vx_i,\vx)(\xi_{,ab}^\phi {\mathcal{B}}^\dagger)(\vy,\vx_j)\delta^D(\vx-\vy) -{\mathcal{B}}(\vx_i,\vx)(\xi_{,ab}^\phi)(\vx,\vy)({\mathcal{B}}^\dagger)(\vy,\vx_j) \\
& - ({\mathcal{B}})(\vx_i,\vy)(\xi_{,ab}^\phi)(\vy,\vx)({\mathcal{B}}^\dagger)(\vx,\vx_i)
+({\mathcal{B}}\xi_{,ab}^\phi)(\vx_i,\vy) ({\mathcal{B}}^\dagger)(\vx,\vx_j)\delta^D(\vx-\vy).
\end{split}
\enq
There are implicit sums over Stokes indices in the above equations.
We introduce further the following notation for the inverse-variance weighted CMB maps $\left (\Res\right)$ and Wiener-filtered CMB $\left (\X_\phi \right)$
\beq
\Res \equiv {\mathcal{B}}^\dagger \Covi \Stdat \quad \quad \textrm{and} \quad \quad
\X_\phi \equiv \xi^\phi {\mathcal{B}}^\dagger \Covi \Stdat = \xi^{\phi}\Res.
\enq
The subscript $\phi$ on the Wiener-filtered CMB is present to emphasize that these are the \textit{lensed} Wiener-filtered delensed CMB maps, in contrast to Eq.~\eqref{eq:XWF}. With this we may write (full indices)
\beq
\left[\frac{\delta{\Cov}}{\delta \alpha_b(\vy)} {\Covi} \Stdat\right](\vx_j) = {\mathcal{B}}(\vx_j,\vy) \X_{,b}(\vy)- \left({\mathcal{B}}\xi^\phi_{,b}\right)(\vx_j,\vy)\Res(\vy)
\enq
The Fisher-like terms contract two such vectors, with an inverse covariance in the middle. When doing the contraction, the operator $\mathcal B^\dagger \Covi \mathcal B \equiv K_\phi $ appear, and derivatives of $\xi$ on the left-hand side gets a minus sign because the first derivative is anti-symmetric. The result is 
\beq
\begin{split}
	H_F^{ab}(\vx,\vy) &= \frac 12\X_{\phi,a}(\vx)K_\phi(\vx,\vy)\X_{\phi,b}(\vy) - \frac 12\Res(\vx)(\xi^\phi_{,a} K_\phi \xi^\phi_{,b})(\vx,\vy) \Res(\vy)\\
	&-\frac12 \X_{\phi,a}(\vx)(K_\phi\xi^\phi_{,b})(\vx,\vy)\Res(\vy) + \frac12\Res(\vx)(\xi^\phi_{,a} K_\phi)(\vx,\vy)\X_{,b}(\vy)
\end{split}
\enq
This matrix is positive definite for any model.
To apply this a vector $v$, we can do as follows
\beq \label{eq:apply2dat}
\begin{split}
\int d^2y \:H_F^{ab}(\vx,\vy)v_b(\vy) =  \X_{\phi,a}(\vx)(V -W)(\vx) + \Res(\vx)\left(\xi_{,a} (V -W)\right)(\vx)
\end{split}
\enq
with
\beq
V = K_\phi\:\X_{\phi,b}\: v^b \quad \quad W = K_\phi \:\xi^{\phi}_{,b}\:\Res \:v^b.
\enq
There is thus only one set of maps to inverse filter, the difference between $(\xi_{,b} \Res)(\vx) v^b(\vx)$ and $(\xi_{,b}(\Res v^b))(\vx)$. There remains the term $H_{\rm det}$, which comes from the second derivative of the covariance in Eq.~\eqref{eq:Hdat}. It is also a contraction of these types of maps. Explicitly,
\beq\
H_{\det}^{ab}(\vx,\vy) = \Res(\vx)\xi_{,ab}^\phi(\vx,\vy) \Res(\vy) - \delta^{D}(\vx-\vy)\Res(\vx)\X_{,ab}(\vx),
\enq
which is easily applied to input vectors. The full $H$ matrix, after rescaling by a simple isotropic approximation, is close enough to unity that its inverse can be applied to a vector with conjugate gradient inversion without much difficulty.

Finally, we reproduce for completeness the expression for the mean-response $R_L$ used in several places in this work, defined above in Eq.~\eqref{eq:mfresp}. Again, we state the result $R^{ab}(\vx, \vy)$ for the two components of the displacement field, defined as
\beq
R^{ab}(\vx, \vy) \equiv  \left. \frac{\delta^2\frac 12 \ln \det \Cov}{\delta \alpha_a(\vx)\delta \alpha_b(\vx)}\right|_{\phi \equiv 0}.
\enq
Since this is evaluated for vanishing deflection, the matrix is only a function of $\vr = \vx - \vy$. The result follows straightforwardly from the covariance first and second variations, Eqs.~\eqref{eq:cov1stvar} and~\eqref{eq:cov2ndvar}:
\beq
\label{eq:MFpred}
-R^{ab}(\vr) =\left(\xi^0_{,a}  K_0  \right)(\vr)\left( \xi^0_{,b}  K_0 \right)(\vr)+  K_0(\vr) (\xi^0_{,a} K_0 \xi^0_{b})(\vr)  -  K_0(\vr)\xi^0_{ab}(\vr) +  \delta^D(\vr)  ( K_0 \xi^0_{,ab})(\vr).
\enq
A contraction on Stokes indices is implicit in this equation.
Contracting $R^{ab}$ after Fourier transformation with $\bL_a \:\bL_b$ gives the lensing potential response $R_L$. All terms are easily calculated with 2-dimensional FFT methods.
\twocolumngrid
\end{document}